# The Nonlinear Theory of FEL

K.S. Badikyan

National University of Architecture and Construction, Yerevan, Armenia

badikyan.kar@gmail.com

**Abstract**

The nonlinear theory of relyativistic strophotron is developed. Classical equations of motion are averaged over fast oscillations. The slow motion phase and saturation parameter are found different from usual undulator oscillation parameters. In the strong field approximation the analytical expression of gain is found on higher harmonics of main resonance frequency.

## 1. Equations of Motion

The relativistic strophotron FEL [1-4] is a system in which electrons move along the parabolic one dimensional potential trough whose vector potential is given by

$$A_z = gx^2/2 \qquad (1)$$

Where x is one of the transverse coordinates, $g$ is the magnetic field gradient. The wave amplified to propagate along the axis of the trough (z - axis). The classical equations of electron motion can be reduced to the form

$$\dot{w} = -\frac{eE_0}{\varepsilon_0}\cos\varphi, \qquad (2)$$

$$(1+w)\dot{\varphi} = \dot{\varphi}(t_0) - \frac{1}{2}\Omega^2\omega\left(x^2 - x_0^2\right), \qquad (3)$$

$$\ddot{x} + \Omega^2 x = -w\ddot{x} - \frac{eE_0}{\varepsilon_0}\left(\frac{\dot{\varphi}}{\omega} - x^2\right)\cos\varphi$$
$$+ \frac{\Omega^2 x}{1+w}\left[\frac{\dot{\varphi}(t_0)}{\omega} - \frac{\Omega^2}{2}\left(x^2 - x_0^2\right)\right], \qquad (4)$$

Where $\mathbf{E}_0$ *parallel Ox* and $\omega$ are the electric field strength and frequency of the wave c=1, e is the electron charge,

$$\varphi = \omega(t-z), \quad w = \frac{\varepsilon - \varepsilon_0}{\varepsilon_0} \qquad (5)$$

Is the time - independent electron energy, the index "0" indicates the initial value of the variables at the time $t = t_0$ when the electron enters the system

$$\Omega = (ge/\varepsilon_0)^{1/2} \qquad (6)$$

is a frequency of strophotron oscillations [1-5].

The terms in of Eq. (4) determine either small corrections to the frequency $\Omega$ (6) or small anharmonicity of oscillations. Here these small corrections will be ignored, and all the mentioned terms in Eq.(4) will be dropped out. In this (harmonic) approximation the field-free $E_0 = 0$ solution of Eq.(4) is given by $x^{(0)} = a_0 \cos[\Omega(t-t_0) + \theta_0]$, where $a_0 = (x_0^2 + \dot{x}_0^2/\Omega^2)^{1/2}$ is the field free amplitude of strophotron oscillations, $\theta_0 = \text{arctg}\, \alpha / x_0 \Omega$, $x_0 = x(t = t_0)$, $\dot{x}_0 = \alpha$, $\alpha \equiv \alpha_x$ is the angle in the plane (x,z) under which the electron enters the system. The first corrections to $x^{(0)}(t)$ ($\propto E_0$) have been found in Ref. [1] and in this approximation x(t) can be written in the form

$$x(t) = a(t) \cos[\Omega(t-t_0) + \theta(t)] \qquad (7)$$

Where $a(t)$ and $\theta(t)$ are some slowly varying functions (amplitude and phase).

$$|\dot{a}/a|, \quad |\dot{\theta}| \ll \Omega. \qquad (8)$$

We will assume that in the general case the expression (7) for x(t) and the condition (8) hold for any values of $E_0$. This assumption about slow variation of $a(t)$ and $\theta(t)$ is justified if

$$|w| \ll 1, \quad a\Omega \ll 1 \qquad (9)$$

i.e. if the total change of electron energy is small in comparison with $\varepsilon_0$ and if the amplitude of oscillations "a" is small in comparison with their period along the z-axis $\lambda_0 = 2\pi/\Omega$. The letter condition means, in particular $\alpha \ll 1$ and $d_e \Omega \ll 1$, where $d_e$ is the diameter of the beam in the (x,z) plane ($x_0 \leq d_e$). We will assume also that $\gamma = \varepsilon_0/m \gg 1$. Under these conditions Eq. (2) gives $|\dot{\varphi}/\omega| \ll 1$. Besides, it is possible to show that the change of amplitude $\delta a \equiv a(t) - a_0$ is small in comparison $a_0$, $|\delta a| \ll a_0$. The condition $a\Omega \ll 1$ (9) justifies also the mentioned above possibility to neglect all the terms in the square brackets of Eq. (4).

The relative emitted energy $w$ contains both fast and slow varying parts $w_{fast}$ and $w_{slow}$. The equations of motion (4) can be averaged over fast period oscillations $2\pi \Omega^{-1}$ to give equation for the slowly varying functions $w_{sl}, \theta$ and $\delta a$ only.

A procedure of separation of fast and slow motions does work well enough only near resonances.

The strophotron FEL is characterized by its resonance frequency

$$\omega_{res} = \frac{\Omega}{\frac{1}{2\gamma^2} + \frac{a_0^2 \Omega^2}{4}}. \tag{10}$$

Amplification can occur near odd harmonics of $\omega_{res}$, $\omega \approx (2s+1)\omega_{res}$, $s = 0, \pm 1, \pm 2, ...$ [5]. We will consider the case $s \gg 1$ assuming the detuning from the (2s+1)-st harmonic

$$\Delta_s = \omega \left( \frac{1}{2\gamma^2} + \frac{a_0^2 \Omega^2}{4} \right) - (2s+1)\Omega \tag{11}$$

Is small $\Delta_s \ll \Omega$.

Under these conditions the equations for $w_{sl}, \theta$ and $\delta a$ can be found from Eq. (2), (3) and (4)

$$\dot{w}_{sl} = -\frac{eE_0 \Omega a_0}{2\varepsilon_0} [J_s(Z) - J_{s+1}(Z)] \sin\psi \tag{12}$$

$$\dot{\theta} = -\frac{\Omega}{2} w_{sl}, \tag{13}$$

$$\delta\dot{a} = -\frac{eE_0}{4\gamma^2 \varepsilon_0 \Omega} [J_s(Z) - J_{s+1}(Z)] \sin\psi, \tag{14}$$

Where $w_{sl}$ is the slow varying part of $w$, $Z = a_0^2 \Omega \omega / 8$,

$$\psi = \omega t_0 + \Delta_s (t - t_0) - (2s+1)\theta + Z \sin 2\theta - \frac{\Omega^2 \omega a_0}{2} \int_{t_0}^{t} \delta a \, dt - \omega \left( \frac{1}{2\gamma^2} + \frac{a_0^2 \Omega^2}{4} \right) \int_{t_0}^{t} w_{sl} \, dt \tag{15}$$

Is the phase of slow motion (resonance) motion of electron. The phase $\psi$ is determined not only by detuning $\Delta_s$ and the phase $\theta$ of the transverse electron coordinate x(t) (7), but also by the change of amplitude of x(t) (7) $\delta a(t)$ and by the change of electron energy w(t).

Combining Eqs. (12), (13) and (14) we find equation for the phase $\psi$ (15), which can be written as the usual pendulum equation [1,5]

$$\frac{d^2\psi}{d\mu^2} = \sin\psi \qquad (16)$$

With the initial conditions

$$\psi(\mu=0) = \omega t_0, \quad \frac{d\psi}{d\mu}(\mu=0) = \frac{\Delta_\varepsilon}{\Delta_m}, \qquad (17)$$

Where the dimensionless time (or the saturation parameter) $\mu$ is given by

$$\mu = t - t_0 - \left\{\frac{eE_0\omega\Omega a_0}{4\varepsilon_0}\left(\frac{1}{2\gamma^2} + \frac{a_0^2\Omega^2}{4}\right)[J_s(Z) - J_{s+1}(Z)]\right\}^{1/2}, \qquad (18)$$

$\Delta_\varepsilon = \varepsilon - \varepsilon_{res}$, $\varepsilon_{res}$ is that value of the initial electron energy at which $\Delta_s = 0$ (10)

$$\Delta_m = \frac{4\varepsilon}{\omega}\tilde{\Delta}_m \Bigg/ \left(-\frac{3}{\gamma^2} + \frac{\alpha^2 + x_0^2\Omega^2}{4}\right),$$
$$\tilde{\Delta}_m = \mu/t. \qquad (19)$$

## 2. Gain

The emitted energy $\Delta\varepsilon = \varepsilon_0 w_{sl}$ is given by Eq. (12) or by the following equivalent equation

$$\Delta\varepsilon = \frac{2\varepsilon}{\omega\left(\frac{1}{\gamma^2} + \frac{a_0^2\Omega^2}{4}\right)}\left(\Delta_s - \tilde{\Delta}_m \frac{d\psi}{d\mu}\right). \qquad (20)$$

In the weak field approximation Eq. (16) solved by the iteration method to give results coinciding with the results of [1,5].

In the strong-field assumptions $\mu \gg 1$, $\tilde{\Delta}_m \gg |\Delta_s|$ the phase $\psi$ obeying the pendulum equation and averaged over its initial value $\psi_0 = \psi(t=t_0)$ and the energy $\overline{\Delta\varepsilon}$ emitted by a single electron are known analytically [1].

The strong field characteristic detuning $\tilde{\Delta}_m$ determines the order of magnitude of the detuning $\Delta_s$ at which the emitted energy $\overline{\Delta\varepsilon}(\Delta_s)$ achieves its maximum. The maximum emitted energy $\overline{\Delta\varepsilon}_{max}$ in the strong field limit has the order of magnitude

$$\Delta\tilde{\bar{\varepsilon}}_{max} \sim \Delta\bar{\varepsilon}(\Delta_s \sim \tilde{\Delta}_m). \tag{21}$$

Hence, as a whole the function $\overline{\Delta\varepsilon}(\Delta_s)$ in the strong field asymptotics $\mu \gg 1$ can be written as

$$\Delta\bar{\varepsilon} = \frac{2\varepsilon\tilde{\Delta}_m}{\omega\left(\frac{1}{\gamma^2} + \frac{a_0^2\Omega^2}{4}\right)}\left\{1 - \frac{2}{\sqrt{\pi\mu}}\sin\left(\mu + \frac{\pi}{4}\right)\right\}F\left(\frac{\Delta_s}{\tilde{\Delta}_m}\right), \tag{22}$$

Where $F(\xi)$ is some dimensionless function, $F(\xi) = -F(-\xi)$, $F(0) = 0$, having its maximum $F(\xi) \sim 1$ at $\xi \sim 1$. In the region $\xi \ll 1$, $F(\xi) \sim \xi$. For any $|\xi| > or < 1$, F cannot be found and written down analytically and requires a numerical calculation [6-9].

Substituting $\omega_s \approx (2s+1)\omega_{res}$ (10) into Eq. (21) we obtain an estimate of efficiency of RS FEL:

$$\eta = \frac{\Delta\bar{\varepsilon}_{max}}{\varepsilon} \sim \frac{\mu}{N_0(2s+1)}, \tag{23}$$

Where $N_0 = \Omega t / 2\pi = L/\lambda_0$ is the number of strophotron oscillations along the length of the system L. When s=0 the estimate (23) coincides with the results well known in the usual undulator FEL [1,5]. When s becomes larger, $\eta$ decreases.

One of the main conclusions of previous papers [1-4] consisted in the prediction of a very strong inhomogeneous broadening connected with a distribution of electrons over the beam's cross section. The result has been derived earlier [1-4] in the weak-field approximation. But, as a matter of fact, this feature of RS FEL in general enough to hold good for any field strength $E_0$. The reason of this inhomogeneous broadening is in a strong dependence of resonance frequency $\omega_{res}$ (10) on the initial transverse electron coordinate $x_0$ (through the amplitude $a_0$). To take into account this inhomogeneous broadening in the strong field limit we must average the emitted energy $\Delta\bar{\varepsilon}$ (22) over $x_0$.

The gain is proportional to the averaged resonance emitted energy $\langle\Delta\bar{\varepsilon}\rangle_{res}$ and has a form

$$G = \frac{8\pi N_e}{E_0^2} \langle \Delta \bar{\varepsilon} \rangle_{res} = \frac{8(\ln 2)^{1/2} \left[ \frac{e\alpha^3}{\varepsilon\omega} (J_s(Z) - J_{s+1}(Z)) \right]^{3/4}}{d_e \Omega E_0^{5/4} \left( \frac{1}{\gamma^2} + \frac{\alpha^2}{4} \right)} \tilde{F}_{res}(\omega) \left\{ 1 - \frac{2}{\sqrt{\pi\mu}} \sin\left( \mu + \frac{\pi}{4} \right) \right\}, \quad (24)$$

Where $N_e$ is the electron number density, $\tilde{F}_{res}(\omega)$ is the form factor of resonance curve.

As a function of $E_0$ the nonlinear gain (24) falls down as $E_0^{-5/4}$ (in contrast with the law $E_0^{-3/2}$ in the usual undulator FEL [1]).

## Discussion

Taking $\omega = \omega_{max} \approx 3\gamma^3 \alpha \Omega$, $s = s_{min}(\omega = \omega_{max}) \approx \frac{3}{8}(\alpha\gamma)^3$ [1] for the saturation parameter $\mu$ we obtain

$$\mu = \left( \frac{\sqrt{3}}{8\pi} \frac{eE_0 \alpha^2 \Omega}{mc} \right)^{1/2} t. \quad (25)$$

A saturation occurs when $\mu > 2$, $E_0 \geq E_{0,sat}$ where the saturation field strength is given by

$$E_{0,sat} = \frac{32\pi mc}{\sqrt{3} e\Omega(\alpha t)^3}. \quad (26)$$

Substituting here $g = 10^4 Oe/cm$, $L = 3m$, $\alpha = 0.7$ we find $E_0 \sim 10^3 V/cm$.

This estimate shows that the gain saturates at rather moderate field $E_0$ and hence the efficiency of the RS FEL is not too high. This result is not surprising because we have considered here an amplification at very high harmonics (s=100). There is an evident way to increase the efficiency constructing a tapered RS FEL, i.e. a system with g=g(z).